\documentclass[pra,aps,showpacs,twocolumn,superscriptaddress,floatfix]{revtex4}
\usepackage{graphicx}
\usepackage[ansinew]{inputenc}
\usepackage{array}
\usepackage{color}
\usepackage{amsmath}
\usepackage{amsxtra}
\usepackage{amstext}
\usepackage{amssymb}
\usepackage{latexsym}
\begin{document}
\title{Deterministic macroscopic quantum superpositions of motion via quadratic optomechanical coupling}
\author{Huatang Tan}
\affiliation{Department of physics, Huazhong Normal University, Wuhan 430079, China}
\affiliation{B2 Institute, Department of Physics and College of Optical Sciences,
The University of Arizona, Tucson, AZ 85721}
\author{F. Bariani}
\affiliation{B2 Institute, Department of Physics and College of Optical Sciences,
The University of Arizona, Tucson, AZ 85721}
\author{Gaoxiang Li}
\affiliation{Department of physics, Huazhong Normal University, Wuhan 430079, China}
\author{P. Meystre}
\affiliation{B2 Institute, Department of Physics and College of Optical Sciences,
The University of Arizona, Tucson, AZ 85721}

\begin{abstract}
We propose a scheme to prepare macroscopic quantum superpositions of motion in optomachanical nano- or micromechanical oscillators quadratically coupled to an intracavity field. The nonlinear optomechanical coupling leads to an effective degenerate three-wave mixing interaction between the mechanical and cavity modes. The quantum superpositions result from the combined effects of the interaction and cavity dissipation. We show analytically and confirm numerically that various deterministic quantum superpositions can be achieved, depending on initial mechanical state. The effect of mechanical damping is also studied in detail via the negativity of the Wigner function. The present scheme can be realized in various optomechanical systems with current technology.
\end{abstract}

\maketitle
\emph{Introduction.}--- While various quantum superposition states have been realized in microscopic physical systems such as atoms and photons \cite{atom, atomphoton}, their realization in macroscopic objects remains a challenging task due to the increasing rate of decoherence induced by coupling to the environment \cite{dec}. Yet, this situation may soon change with recent rapid experimental progress in quantum optomechanics --- the interface between nano or micromechanical oscillators and quantum optical fields~ \cite{opm}. In particular, cooling macroscopic mechanical oscillators to their quantum ground state~\cite{c1,c2,c3,c4} and realizing strong optomechanical coupling \cite{c4,sc1-mb,sc2,sc3} will lay a solid foundation for exploring macroscopic quantum states theoretically and experimentally in macroscopic systems \cite{qzm1, qzm2, rabl, yuxi, cat0, cat1,cat2,cat3,cat4}. Besides their fundamental interest, these states are potential resources in various quantum technologies \cite{qi, qmetr, qbit}.

This paper analyzes theoretically a new scheme to prepare macroscopic quantum superposition states of nano- or micromechanical oscillators coupled quadratically to a light field. In contrast  to existing schemes that rely on conditional quantum measurements~\cite{cat0, cat1,cat2,cat3} or quantum state transfer~\cite{cat4}, this scheme is deterministic. It relies on the combined effects of two-phonon emission and absorption that can be engineered with quadratic optomechanical coupling. Pure mechanical quantum superposition states can be generated for time scales over which mechanical damping remains negligible. This approach could also be applied to other physical systems in which degenerate three-wave mixing can be engineered.
\begin{figure}[t]
\centerline{\scalebox{0.5}{\includegraphics{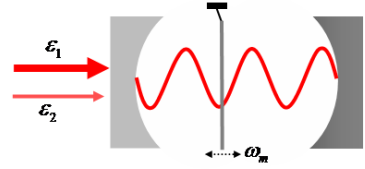}}}
 \caption{Possible membrane-in-the-middle setup for the steady-state generation of macroscopic superpositions, with a pure quadratically coupled to two classical driving fields to generate an effective two-phonon three-wave mixing interaction.}
\end{figure}

\emph{Model.}--- We consider a ``membrane-in-the-middle'' cavity optomechanical system. The mechanical oscillator could be e.g. a suspended dielectric membrane (as shown in Fig. 1) \cite{sc1-mb}, an ultracold atomic cloud \cite{bec}, or a trapped levitated dielectric sphere \cite{sph}. For such geometries either linear or quadratic optomechanical coupling can be achieved by adjusting the equilibrium position of the oscillator in the standing-wave cavity field, and purely quadratic coupling can be obtained by positioning the mechanical oscillator at a node or antinode of the cavity field, as demonstrated in Refs. \cite{sc1-mb, bec}. In our scheme quadratic coupling is exploited to achieve macroscopic quantum superpositions of motion, while linear coupling with a weak probe field could be used to characterize the motional superpositions.

We assume that a cavity mode of resonance frequency $\omega_c$ is driven by two laser fields of frequencies $\omega_j$ and amplitudes $\mathcal{E}_j~(j=1,2)$. In the rotating frame with respect to the frequency $\omega_1$, the Hamiltonian of the optomechanical system reads
\begin{align}
\hat H_0/\hbar&=\delta_c \hat A^\dag \hat A
+\omega_m \hat B^\dag \hat B+g_0\hat A^\dag \hat A(\hat B+\hat B^\dag)^2\nonumber\\
&+(\mathcal{E}_1+\mathcal{E}_2e^{-i\delta_{12}t}) \hat A^\dag+(\mathcal{E}_1^*+\mathcal{E}_2^*e^{i\delta_{12}t}) \hat A,
\label{hamil}
\end{align}
where $\hat A$ and $\hat B$ ($\hat A^\dag$ and $\hat B^\dag$) are the annihilation (creation) operators of the cavity mode and the mechanical mode of frequency $\omega_m$, respectively, the detunings are $\delta_c=\omega_c-\omega_1$ and $\delta_{12}=\omega_2-\omega_1$, and $g_0$ is the single-photon quadratic coupling frequency. The last two terms in Eq.~(\ref{hamil}) describe the pumping of the cavity by the external lasers.

For sufficiently strong classical driving fields we can decompose the field operators $\hat A$ and $\hat B$ as the sum of their expectation values and small quantum fluctuations, $\hat A=\alpha+ \hat a$ and $\hat B=\beta+ \hat b$. The classical amplitudes $\alpha\equiv\langle\hat A\rangle$ and $\beta\equiv\langle\hat B\rangle$ are determined by the equations
\begin{subequations}
\begin{align}
\dot{\alpha}&=-(\kappa_c+i\delta_c)\alpha-ig_0\alpha(\beta+\beta^*)^2
-i(\mathcal{E}_1+\mathcal{E}_2e^{-i\delta_{12}t}),\label{sm1}\\
\dot{\beta}&=-(\gamma_m+i\omega_m)\beta-2ig_0|\alpha|^2(\beta+\beta^*).
\end{align}
\end{subequations}
where $\kappa$ is the cavity dissipation rate and $\gamma_m$ the mechanical damping rate. For $|\mathcal{E}_1|\gg|\mathcal{E}_2|$ this latter amplitude can be neglected in Eq.~(\ref{sm1}) and the steady-state cavity amplitude is approximately $\alpha_s\simeq \mathcal{E}_1/(i\kappa_c-\Delta_c)$, where $\Delta_c=\delta_c-g_0(\beta_s+\beta_s^*)^2$. That amplitude can be taken to be real by a proper choice of the phase of $\mathcal{E}_1$. The steady-state amplitude $\beta_s$ is determined by
\begin{align}
(4g_0\omega_m|\alpha_s|^2-\omega_m^2-\gamma_m^2)(\beta_s+\beta_s^*)=0.
\end{align}
Unless the first term in parenthesis is equal to zero, and in particular for $g_0< 0$, a situation corresponding to the oscillator at a minimum of the standing-wave intracavity field, this gives $\beta_s=0$.

In the following we consider the specific situation where the cavity is resonantly driven by the weak laser field and by the second red-detuned sideband of the strong laser field,
\begin{subequations}
\begin{gather}
\delta_c=\omega_c-\omega_1=2\omega_m, \\
\omega_2=\omega_c.
\end{gather}
\end{subequations}
Substituting the expressions for the operators $\hat A$ and $\hat B$ into the Hamiltonian~(\ref{hamil}), considering a mechanical frequency $\omega_m\gg g\equiv g_0\alpha_s$, and neglecting  the rapidly oscillating terms, the resonant optomechanical Hamiltonian becomes then approximately
\begin{align}
\hat H =\hbar g\hat a\hat b^{\dag2}+\hbar \mathcal{E}_2 \hat a+{\rm h.c.},
\label{hami2}
\end{align}
where $g$ can be controlled by adjusting the laser amplitude $|\mathcal{E}_1|$. This Hamiltonian can be interpreted as describing the degenerate three-wave mixing of optical and acoustic waves, whereby a photon is absorbed from (or emitted into) the pump laser of classical amplitude $\mathcal{E}_2$ and two phonons are simultaneously created (or annihilated). Due to the correlated nature of two-phonon emission, quantum features such as phononic squeezed states can be achieved in such a process~\cite{phnsqz}. Here we take advantage instead of this interaction to explore the generation of macroscopic quantum superpositions of mechanical motion.

When taking into account cavity dissipation and mechanical damping, the density operator $\rho$ of the full optomechanical system is governed by the master equation
\begin{eqnarray}
\label{master1}
\dot {\rho}(t)&=&\frac{1}{i\hbar}[\hat H, \rho]+\mathcal{\hat L}_\kappa\rho+\mathcal{\hat L}_{\gamma_m}\rho,
\end{eqnarray}
with
\begin{eqnarray}
\mathcal{\hat L}_\kappa\rho&=&\kappa(2\hat a\rho \hat a^\dag- \hat a^\dag \hat a\rho-\rho \hat a^\dag \hat a),\nonumber\\
\mathcal{\hat L}_{\gamma_m}\rho&=&\gamma_m(n_{\rm th}+1)(2\hat b\rho \hat b^\dag- \hat b^\dag \hat b\rho- \rho \hat b^\dag \hat b)\nonumber\\
&+&\gamma_mn_{\rm th}(2\hat b^\dag\rho \hat b- \hat b \hat b^\dag\rho- \rho \hat b \hat b^\dag ),\nonumber
\end{eqnarray}
where  $\mathcal{\hat L}_k\rho$ accounts for the optical cavity losses, $\mathcal{\hat L}_{\gamma_m}\rho$ describes the mechanical damping in its thermal environment, and the mean thermal phonon number is $n_{\rm th}=(e^{\hbar\omega_m/k_BT}-1)^{-1}$, with $k_B$ the Boltzmann constant and $T$ the temperature. Equation~(\ref{master1}) is our starting point to investigate the generation of quantum superposition states of the mechanical oscillator.

\emph{Dissipation-induced superpositions.}---We proceed by first neglecting mechanical damping, $\gamma_m=0$, in which case it is not difficult to find that with the aid of cavity dissipation the full system evolves toward the dark state
\begin{align}
|\psi_{\rm d} \rangle=|0_a\rangle|\psi_{b,s}\rangle,
\label{dark}
\end{align}
where the mechanical steady state $|\psi_{b,s}\rangle$ satisfies the equation
\begin{equation}
(\hat b^2+\mathcal{E}_2/g) |\psi_{b,s}\rangle=0,
\label{eigensta}
\end{equation}
or equivalently $\hat b^2|\psi_{b,s}\rangle=\mathcal{E}_0|\psi_{b,s}\rangle$, with $\mathcal{E}_0=-\mathcal{E}_2/g$. It is readily verified that the state~(\ref{dark}) then satisfies the dark state conditions $\hat H|\psi_{\rm d}\rangle=0$ and $\mathcal{\hat L_\kappa}(|\psi_{\rm d}\rangle \langle\psi_{\rm d}|)=0$.

\begin{figure}[t]
\centerline{\scalebox{0.65}{\includegraphics{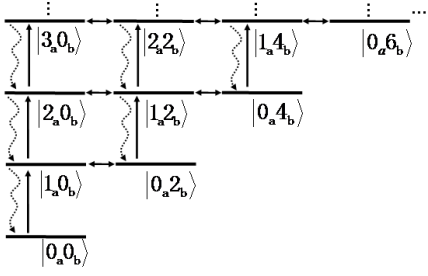}}}
 \caption{Schematic of the processes involved in the realization of the dark state $|\psi_{b,s}\rangle_0$  in  case the initial mechanical state contain only even phonon numbers. The straight vertical arrows describe the optical driving of the resonator, the wavy vertical arrows account for optical dissipation, and the horizontal arrows show the two-phonon transitions in the mechanical oscillator.}
 \label{fig2}
\end{figure}

By expressing $|\psi_{b,s}\rangle$ in the Fock basis $\{|n_b\rangle\}$ as $|\psi_{b,s}\rangle=\sum_{n_b}c_{n_b}|n_b\rangle$, it follows from Eq.~(\ref{eigensta}) that
\begin{align}
\label{ameq}
c_{n_b+2}=\mathcal{E}_0c_{n_b}/\sqrt{(n_b+1)(n_b+2)}.
\end{align}
That the probability amplitudes $c_{n_b}$ with even (or odd) $n_b$ are only related to those with even (or odd) phonon numbers in the absence of mechanical dissipation follows from the two-phonon nature of the interaction. For a mechanical mode initially in an even Fock state we then find from Eq.~(\ref{ameq}) that
\begin{eqnarray}
|\psi_{b,s}\rangle_0&=&\frac{1}{\sqrt{\cosh |\mathcal{E}_0}|}\sum_{n_b=0}^\infty \frac{\mathcal{E}_0^{n_b}}{\sqrt{(2n_b)!}}|2n_b\rangle\nonumber\\
&=&\mathcal{N}_e^{-1/2}\left (|\beta\rangle+|-\beta\rangle\right ),
\end{eqnarray}
which is an even coherent state, with $\beta=\sqrt{\mathcal{E}_0}$ and the normalization $\mathcal{N}_e=2[1+\exp(-2|\beta|^2)]$. Likewise, when starting from an initial odd Fock state, the mechanical mode decays to the odd coherent state
\begin{eqnarray}
|\psi_{b,s}\rangle_1&=&\frac{1}{\sqrt{\cosh |\mathcal{E}_0}|}\sum_{n_b=0}^\infty \frac{\mathcal{E}_0^{n_b}}{\sqrt{(2n_b+1)!}}|2n_b+1\rangle\nonumber\\
&=&\mathcal{N}_o^{-1/2} \left( |\beta\rangle-|-\beta\rangle \right ),
\end{eqnarray}
with $\mathcal{N}_o=2[1-\exp(-2|\beta|^2)]$. As schematically illustrated in Fig.~\ref{fig2}, these steady-state superpositions result from the combined effects of cavity driving and dissipation and two-phonon emission and absorption. Such superpositions was proposed in a trapped cold ion via qubit-phonon coupling \cite{evenodd}.

It is generally much easier to prepare a mechanical vacuum or coherent state than a Fock state with an odd or even initial phonon number. In  that case the mechanical oscillator will evolve toward the mixed state
\begin{eqnarray}
\rho_{b,s}&=&P_{00}|\psi_{b,s}\rangle_{00}\langle \psi_{b,s}|+P_{11}|\psi_{b,s}\rangle_{11}\langle \psi_{b,s}|\nonumber\\
&+& \left ( P_{01}|\psi_{b,s}\rangle_{01}\langle\psi_{b,s}| + {\rm h.c.}\right ),
\label{genesta}
\end{eqnarray}
where $P_{00}=\sum_{n_b}\langle 2n_b|\rho_b(0)|2n_b\rangle$, $P_{11}=\sum_{n_b}\langle 2n_b+1|\rho_b(0)|2n_b+1\rangle$, and $\rho_b(0)$ is the initial state of the mechanical oscillator.

The purity of that state is characterized by the cross-term probability  $P_{01}$, which depends in general on the initial coherence of the mechanics as well as on the system parameters. Assuming that $\mathcal{E}_2=0$ we can obtain an analytical estimate of its value in the bad cavity limit, in which case one can eliminate adiabatically the cavity mode, giving the reduced master equation for the oscillator
\begin{equation}
\label{master2}
\frac{d}{dt}\rho_b(t) =
\gamma_2 (2\hat b^2\rho_b \hat b^{\dag2}- \hat b^{\dag2} \hat b^2\rho_b-\rho_b \hat b^{\dag2} \hat b^2).
\end{equation}
This equation describes an effective dissipation mechanism via two-phonon absorption at the rate $\gamma_2=g^2/\kappa$. We have from Eq.~(\ref{genesta}) that the initial state $\rho_b(0)=|\beta_0\rangle\langle\beta_0|$, with $\beta_0=|\beta_0|\exp(i\theta_0)$, will decay to the steady state
\begin{eqnarray}
\rho_{b,s}=P_{00}|0_b\rangle\langle0_b|+P_{11}|1_b\rangle\langle1_b| +(P_{01}|0_b\rangle\langle1_b|+ \rm{h.c.}),\nonumber\\
\end{eqnarray}
where  $P_{00}=1-P_{11}=[1+\exp(-2|\beta_0|^2)]/2$ and $P_{01}=|\beta_0|e^{-|\beta_0|^2-i\theta_0}I_0(|\beta_0|^2)$, where $I_0(x)$ is the modified Bessel function of the first kind \cite{coh}.
For the relatively large initial coherent amplitude $|\beta_0|=1.7$, the probability amplitudes $P_{00}\simeq P_{11}\simeq0.5$ and $|P_{01}|\simeq0.42$. That is, the state reached by the oscillator is approximately the quantum superposition
\begin{eqnarray}
|\psi_{b,s}\rangle_2\simeq (|0_b\rangle+e^{i\theta_0}|1_b\rangle)/\sqrt{2}.
\end{eqnarray}
\begin{figure}[t]
\centerline{\scalebox{0.8}{\includegraphics{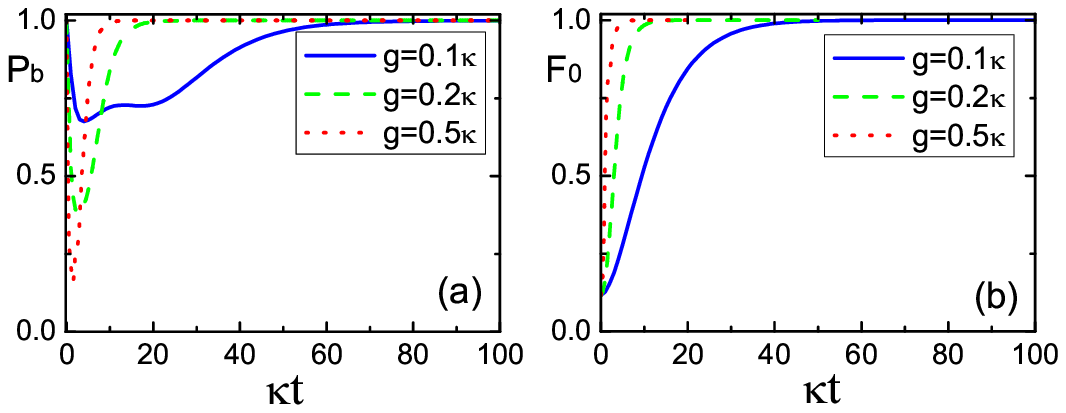}}}
\centerline{\scalebox{0.4}{\includegraphics{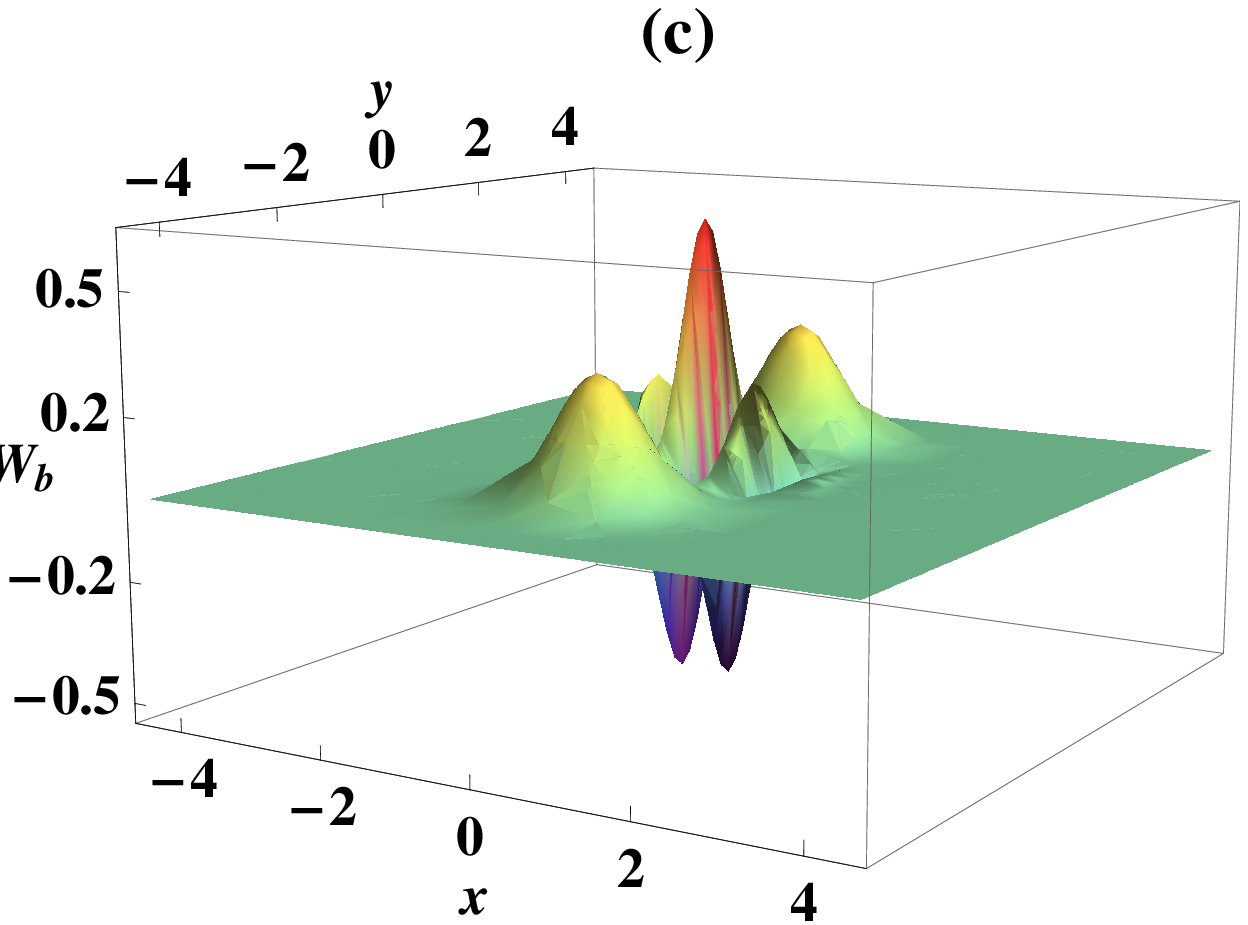}}}
 \caption{Time evolution of (a) the purity $P_b$ and (b) the fidelity $F_0=\sqrt{_0\langle \psi_{b,s}|\rho_b|\psi_{b,s}\rangle_0}$ for $\beta=\sqrt{5}$  and initial vacua of both the cavity and the mechanical modes. Here the pumping amplitude $\mathcal{E}_2=5g$ and the mechanical damping $\gamma_m=0$. (c) Wigner function $W_b$ of the mechanical state in the long-time regime. Time in units of $1/\kappa$.}
 \label{purecat}
\end{figure}

\emph{Numerical results.}---To confirm these approximate analytical results and evaluate the effects of thermal fluctuations, we solved numerically the full master equation (\ref{master1}), expanding the density matrix $\rho$ on the Fock basis $\{|n_a, n_b\rangle\}$ as
\begin{align}
\rho=\sum_{m_a, m_b, n_a, n_b}\rho_{m_a, m_b, n_a, n_b}|n_a, n_b\rangle\langle m_a, m_b|.
\end{align}
and truncating the system to a finite Fock space large enough to avoid boundary issues. We used the Wigner function~\cite{barnett}
\begin{align}
W_b(\zeta)=\frac{1}{\pi^2}\int d^2\eta {\rm Tr}(e^{\eta\hat b^\dag-\eta^*\hat b}\rho_b) e^{\zeta\eta^*-\zeta^*\eta},
\end{align}
where $\rho_b\equiv\rm Tr_a (\rho)$ is the reduced mechanical density matrix, to reveal the quantum features of the state of the oscillator: negative Wigner functions are indicative of highly nonclassical superposition states. In addition, the purity of the mechanical states is characterized by the quantity $P_b={\rm Tr}(\rho_b^2)$.
\begin{figure}
\centerline{\scalebox{0.38}{\includegraphics{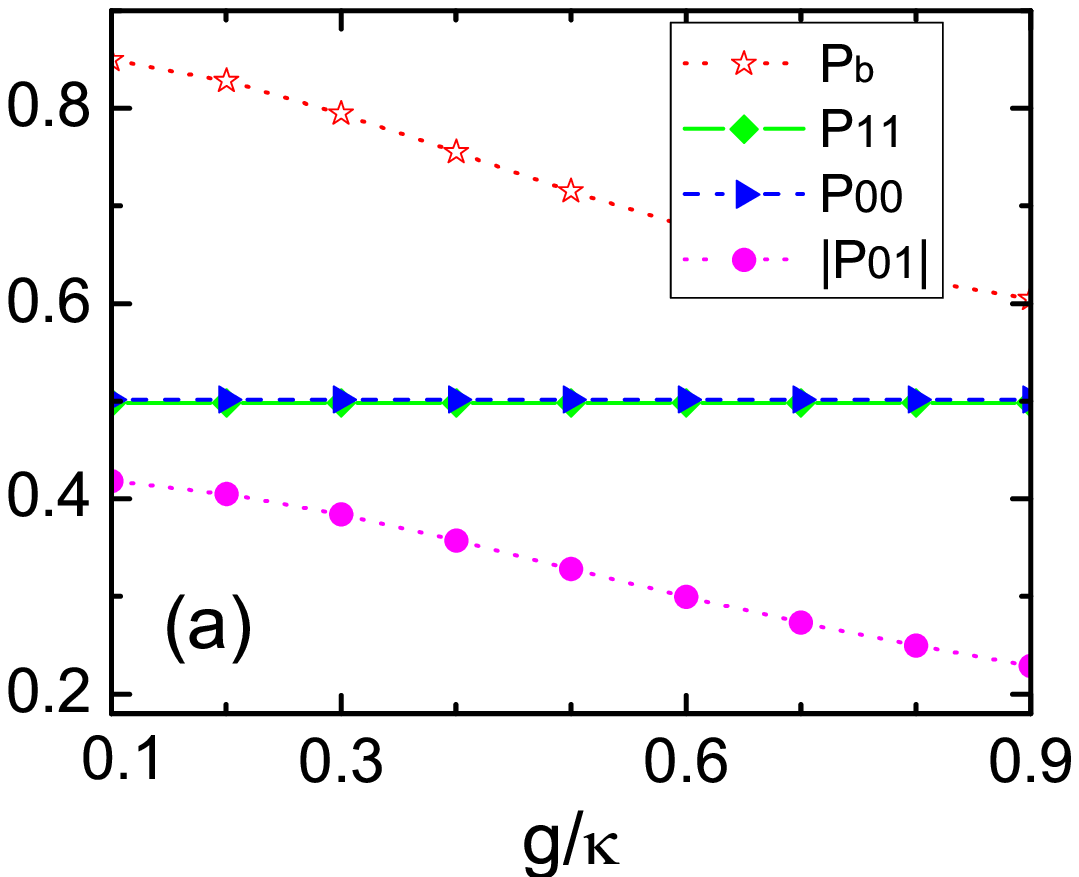}}\scalebox{0.38}{\includegraphics{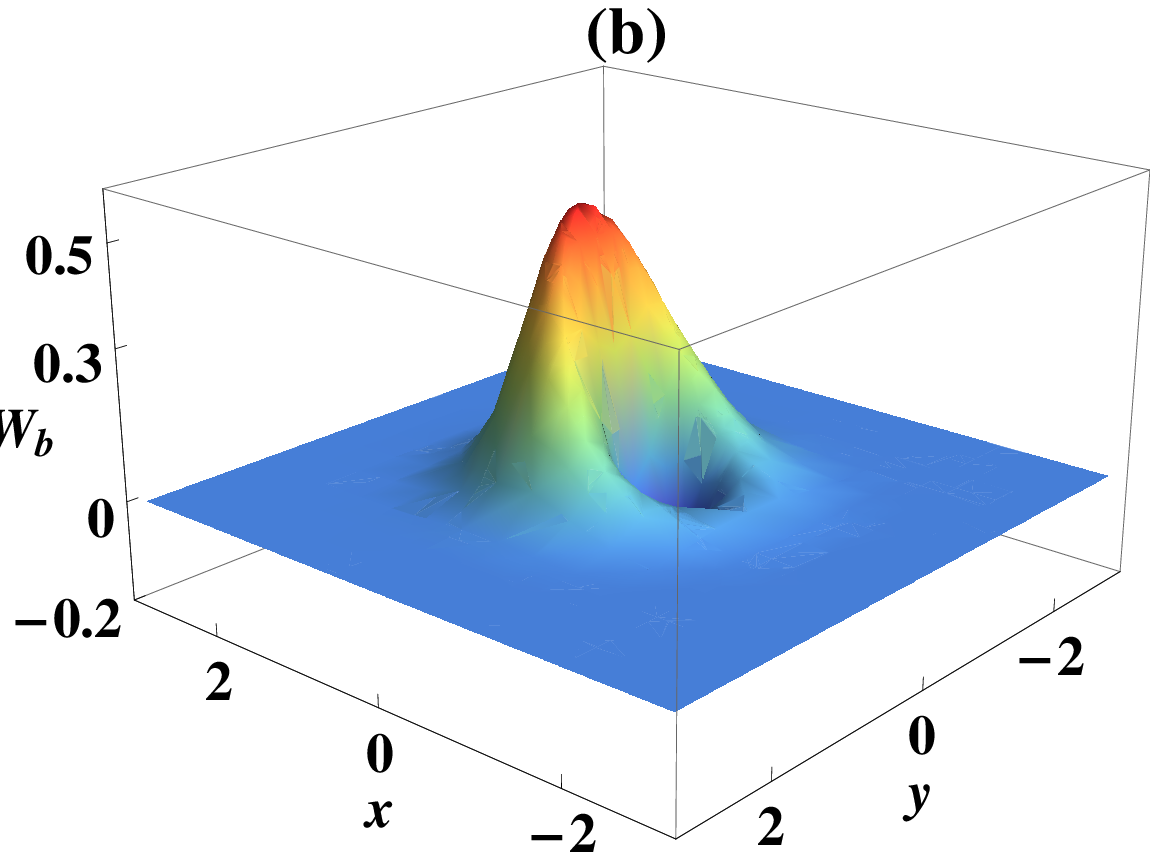}}}
 \caption{(a) Dependence of the purity $P_b$, the probability amplitudes $P_{00}$ and  $P_{11}$, and the coherence $|P_{01}|$ of the steady mechanical states on the ratio $g/\kappa$ for initial cavity vacuum and mechanical coherent state $|\beta_0\rangle$ with $|\beta_0|=1.7$. Here the pumping amplitude $\mathcal{E}_2=0$ and the mechanical damping $\gamma_m=0$. (b) Wigner function $W_b$ of the steady mechanical state for $g=0.1\kappa$.}
 \label{puregate}
\end{figure}

Fig.s~3 (a) and 3 (b) show the time evolution of the purity $P_b$ and of the fidelity $F_0=\sqrt{_0\langle \psi_{\rm b,s}|\rho_b|\psi_{\rm b,s}\rangle_0}$ of the mechanical state compared to the ``ideal'' dark state $|\psi_{\rm b,s}\rangle_0$ for several values of the coupling strength $g$ and in the absence of mechanical damping, $\gamma_m=0$. Both of the cavity and mechanical modes start from vacua and the pumping amplitude $\mathcal{E}_2=5g$. As expected, the mechanical oscillator evolves asymptotically from an initial vacuum to the pure even coherent state $|\psi_{\rm b,s}\rangle_0$, the steady-state superposition exhibiting highly nonclassical features demonstrated by the negativity of its Wigner function $W_b$, see Fig.~3 (c).

The steady-state populations $P_{00}$ and $P_{11}$, the coherence $|P_{01}|$, and the purity $P_b$ of the mechanical state when the mechanical oscillator starts from a coherent state $|\beta_0\rangle$ with $|\beta_0|=1.7$ and $\mathcal{E}_2=0$ are shown in Fig.~4 (a). We find $P_{00}\simeq P_{11}\simeq 0.5$, independently of the system parameters. However, $|P_{01}|$ decreases, from the maximum value about $0.42$ for $g=0.1\kappa$, as the coupling $g$ is increased and the system departs from the bad-cavity limit. Likewise, the purity increases with decreasing $g/\kappa$, in good agreement with the analytical results.

\begin{figure}
\centerline{\scalebox{0.8}{\includegraphics{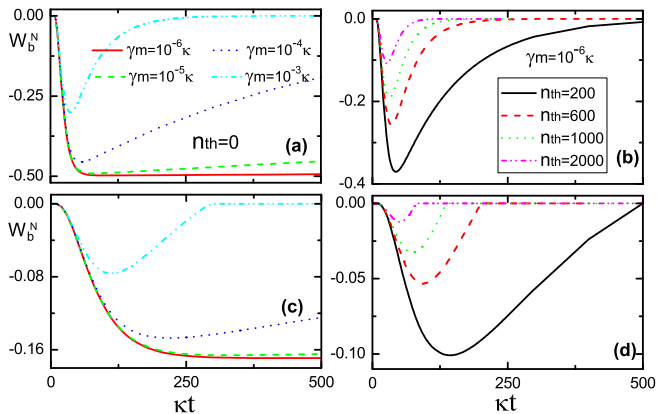}}}
\caption{(a) and (b) Time evolution of the minimal negativity $W_b^N$ of the Wigner function for cavity and mechanical modes initially in a vacuum. Here the pumping amplitude $\mathcal{E}_2=5g$ and coupling $g=0.1\kappa$. (c) and (d) are the same as (a) and (b), respectively, except that the mechanical mode is initially in the coherent state $|\beta_0\rangle$ with $\beta_0=1.7$ and $\mathcal{E}_2=0$.}
\end{figure}

The effect of the mechanical damping $\gamma_m$ on the mechanical superposition is illustrated in Fig. 5. The dynamics is now largely determined by two characteristic times: the first one is the inverse optomechanical frequency $g^{-1}$, and the second the much longer time $\gamma_m^{-1}$ required for the mechanics to reach a steady state. As would be expected the negativity of the Wigner function disappears over that longer time scale. Still, for mechanical oscillators with high quality factor $Q=\omega_m/\gamma_m$, such as e.g. the system reported in Ref.~\cite{sc1-mb}, with $\omega_m/2\pi=10^6~\rm Hz$, $\gamma_m/2\pi=0.1~\rm Hz$, and $\kappa/2\pi=10^5~\rm Hz$,  highly nonclassical quantum superpositions of high purity can be still achieved in the short-time regime for temperatures as high as $T\simeq10~\rm mK$ (with a corresponding thermal phonon number of $n_{\rm th}=200$), as illustrated in Fig.~5 (b) and (d).

\emph{Discussion and conclusions.}---In summary we have proposed a deterministic scheme to generate macroscopic quantum superposition states of a micromechanical oscillators by quadratic cavity optomechanics. These quantum superpositions result from the combined effects of optical damping and the degenerate three-wave mixing interaction between optical and acoustic waves resulting from quadratic optomechanical coupling. We showed analytically and verified numerically that the mechanical system can approach asymptotically a variety of quantum superpositions, depending on the initial mechanical state and pumping condition. Pure even and odd coherent superposition states can be obtained provided that the mechanical damping is neglected. It is also revealed that in the presence of the mechanical damping, the highly nonclassical superpositions can still be achieved in the transient regime for the realistic system parameters.

This proposed scheme can be realized in a number of state-of-the-art optomechanical systems. It could also be used to produce quantum superpositions in other systems where degenerate three-wave mixing can be engineered, such as microwave photons in circuit-QED configurations~\cite{crqed}.

The experimental verification and characterization of these quantum superposition states could be carried out by exciting another cavity mode linearly coupled to the mechanical oscillator with a red-tuned probe laser. In the weak coupling regime, quantum-state transfer between the cavity and the mechanical modes can then take place via a beam-splitter interaction~\cite{om}. Once the mechanical state is completely mapped onto the cavity field, one can reconstruct its Wigner function by quantum homodyne tomography via detecting the output field, see e.g.~Ref.~\cite{qtom}. Future work will consider non-degenerate three-wave mixing in a cavity with two mechanical oscillators, with the goal to prepare non-Gaussian entangled states such as N00N states \cite{noon}.

Helpful discussions with Lukas Buchmann and HyoJun Seok are gratefully acknowledged. This work is supported by the National Natural Science Foundation of China (Grant Nos.~11274134, 11074087, and 61275123),
the National Basic Research Program of China (Grant No. 2012CB921602),
the DARPA QuASAR and ORCHID programs through grants from AFOSR and ARO, the US Army Research Office, and the US National Science Foundation. T.H.T. is also supported by the CSC.

\emph{Note added.}--- After completion of this work, we became aware of the preprint papers, arXiv:1212.4795 by M. J. Everitt, T. P. Spiller, G. J. Milburn, R. D. Wilson, and A. M. Zagoskin; arXiv:1302.1707 by A. Voje, A. Croy, and A. Isacsson, on the generation of mechanical states by two-phonon process.

\end{document}